\documentclass[12pt]{article}

\usepackage{amstext}
\usepackage{latexsym}
\usepackage{amssymb}
\usepackage{amsmath}

\def\R{\mathbb R}

\def\N{\mathbb N}

\newcommand{\eps}{\varepsilon}

\renewcommand{\div}{\nabla\cdot}
\newcommand{\curl}{\nabla\times}

\begin{document}

%
%
%
%
%
%
%
%

\newtheorem{theorem}{Theorem}[section]
\renewcommand{\thetheorem}{\arabic{section}.\arabic{theorem}}
\newtheorem{definition}[theorem]{Definition}
\newtheorem{deflem}[theorem]{Definition and Lemma}
\newtheorem{lemma}[theorem]{Lemma}
\newtheorem{example}[theorem]{Example}
\newtheorem{remark}[theorem]{Remark}
\newtheorem{remarks}[theorem]{Remarks}
\newtheorem{cor}[theorem]{Corollary}
\newtheorem{pro}[theorem]{Proposition}
\newtheorem{proposition}[theorem]{Proposition}

\renewcommand{\theequation}{\thesection.\arabic{equation}}

\title{The Darwin Approximation of the Relativistic Vlasov-Maxwell System}
\author{{\sc Sebastian Bauer\footnote{Supported in parts by
        DFG priority research program SPP 1095}\,\,\,\& Markus Kunze} \\[2ex]
        Universit\"at Essen, FB 6 -- Mathematik, \\
        D\,-\,45117 Essen, Germany \\[1ex]
        \\ \hspace{-4.3em} post-Newtonian approximation}
\date{}
\maketitle
\begin{abstract}\noindent
We study the relativistic Vlasov-Maxwell system which describes large systems of particles
interacting by means of their collectively generated forces. If the speed of light $c$
is considered as a parameter then it is known that in the Newtonian limit $c\to\infty$
the Vlasov-Poisson system is obtained. In this paper we determine the next order
approximate system, which in the case of individual particles usually is called
the Darwin approximation.
\end{abstract}


\setcounter{equation}{0}

\section {Introduction and Main Results}

The relativistic Vlasov-Maxwell system
\begin{equation}\label{RVMC}
   \left\{\begin{array}{lclcrcl}
   \lefteqn{\partial_t f+\hat{v}\cdot\nabla_x f
   +(E+c^{-1}\hat{v}\times B)\cdot\nabla_v f=0,} \\[1ex]
   \;c\curl E & = & -\partial_t B, & & c\curl B  & = & \partial_t E+4\pi j, \\
   \quad\, \div E & = & 4\pi\rho, & \quad & \div B & = & 0, \\[1ex]
   \qquad\quad\rho & := & \int f\,dv, & & j & := & \int\hat{v} f\,dv,
   \end{array}\right.\tag{RVMc}
\end{equation}
describes the time evolution of a single-species system of particles
(with mass and charge normalized to unity) which interact by means of their collectively
generated forces. The distribution of the large number of particles in configuration space
is modelled through the non-negative density function $f(x, v, t)$, depending on position
$x\in\R^3$, momentum $v\in\R^3$, and time $t\in\R$, whereas
\begin{equation}\label{hatv-def}
   \hat{v}=(1+c^{-2} v^2)^{-1/2}v\in\R^3
\end{equation}
is the relativistic velocity associated to $v$. The Lorentz force
$E+c^{-1}\hat{v}\times B$ realizes the coupling of the Maxwell fields
$E(x, t)\in\R^3$ and $B(x, t)\in\R^3$ to the Vlasov equation,
and conversely the density function $f$ enters the field equations via the scalar charge density
$\rho(x, t)$ and the current density $j(x, t)\in\R^3$, which act as source terms
for the Maxwell equations. It is supposed that collisions in the system are sufficiently rare so that they can be neglected. The parameter $c$ denotes
the speed of light, and $\int$ always means $\int_{\R^3}$. At time $t=0$,
the initial data
\[ f(x, v, 0)=f^\circ(x, v),\quad E(x, 0)=E^\circ(x),\quad\mbox{and}\quad
   B(x, 0)=B^\circ(x) \]
are prescribed. In this work we treat the speed of light as a parameter
and study the behavior of the system as $c\to\infty$. Conditions will be establish
under which the solutions of (\ref{RVMC}) converge to a solution of an effective system. We recall that in \cite{schaeffer:86} it has been shown that as $c\to\infty$ the solutions of (\ref{RVMC}) approach a solution of the Vlasov-Poisson system
with the rate ${\cal O}(c^{-1})$; see \cite{asanouk,degond} for similar results
and \cite{lee} for the case of two spatial dimensions. The respective Newtonian limits
of other related systems are derived in \cite{ADR,calee}. It is the goal of this paper
to replace the Vlasov-Poisson system by another effective equation
to achieve higher order convergence and a more precise approximation.
This will lead to an effective system whose solution stays as close
as ${\cal O}(c^{-3})$ to a solution of the full Vlasov-Maxwell system,
if the initial data are matched appropriately. In the context of individual particles, this post-Newtonian order of approximation is usually called the Darwin order,
see \cite{Sp,KS1} and the references therein. Let us also mention that
weak convergence properties of other kinds of Darwin approximations for the Vlasov-Maxwell system were studied in \cite{degondravi,sonnen}. In the present paper
we mainly view the Darwin approximation as a rigorous intermediate step towards the next order, where in analogy to the case of individual particles \cite{KS2} radiation effects are expected to play a role for the first time. Since at the radiation order the corresponding dynamics of the Vlasov-Maxwell system most likely will have to be restricted to a center manifold-like domain in the infinite dimensional space of densities (to avoid ``run-away''-type
solutions \cite{Sp,KS2}), it is clear that several new mathematical difficulties will have to be surrounded in this next step. Then the ultimate goal would be to determine the effective equation for the Vlasov-Maxwell system on the center manifold, which should finally lead to a slightly dissipative Vlasov-like equation,
free of ``run-away'' solutions; see \cite{KR1,KR2} for a model of this equation
and more motivation.

Compared to systems of coupled individual particles, for the Vlasov-Maxwell system
one immediately encounters the problem that so far in general only the existence
of local solutions is known. These solutions are global under additional conditions,
for instance if a suitable a priori bound on the velocities is available;
see the pioneering work \cite{glstr}, and also \cite{klsta,bogopa},
where this result is reproved by different methods.
This means that from the onset we will have to restrict ourselves
to solutions of (\ref{RVMC}) which are defined on some time interval $[0, T]$
that may be very small. On the other hand, in \cite{schaeffer:86}
it has been shown that such a time interval can be found which is uniform in $c\ge 1$,
so it seems reasonable to accept this restriction.

In order to find the desired higher-order effective system, we formally
expand all quantities arising in (\ref{RVMC}) in powers of $c^{-1}$:
\begin{eqnarray*}
   f & = & f_0+c^{-1}f_1+c^{-2}f_2+\ldots, \\
   E & = & E_0+c^{-1}E_1+c^{-2}E_2+\ldots, \\
   B & = & B_0+c^{-1}B_1+c^{-2}B_2+\ldots, \\
   \rho & = & \rho_0+c^{-1}\rho_1+c^{-2}\rho_2+\ldots, \\
   j & = & j_0+c^{-1}j_1+c^{-2}j_2+\ldots,
\end{eqnarray*}
where $\rho_k=\int f_k\,dv$ and $j_k=\int v f_k\,dv$
for $k=0, 1, 2, \ldots$. Moreover, $\hat{v}=v-(c^{-2}/2)v^2v+\ldots$
by (\ref{hatv-def}), where $v^2=|v|^2$. The expansions can be substituted into (\ref{RVMC}),
and comparing coefficients at every order gives a sequence of equations
for these coefficients.

At zeroth order we obtain
\begin{equation}\label{Feldnull}
   \curl E_0=0,\quad\div E_0=4\pi\rho_0,\quad\curl B_0=0,\quad\div B_0=0.
\end{equation}
If we set
\[ B_0=0, \]
then the Vlasov-Poisson system
\begin{equation}\label{VP}
   \left\{\begin{array}{c}\partial_t f_0+v\cdot\nabla_x f_0
   + E_0\cdot\nabla_v f_0=0, \\[1ex]
   E_0(x, t)=-\int |z|^{-2}\bar{z}\,\rho_0(x+z, t)\,dz, \\[1ex]
   \rho_0=\int f_0\,dv, \\[1ex]
   f_0(x, v, 0)=f^\circ(x, v),\end{array}\right.\tag{VP}
\end{equation}
is found, with $\bar{z}=|z|^{-1}z$.

Next we consider the equations at first order in $c^{-1}$. Here
\begin{equation}\label{Feldeins}
   \curl E_1=-\partial_t B_0=0,\quad\div E_1=4\pi\rho_1,
   \quad\curl B_1=\partial_t E_0+4\pi j_0,\quad\div B_1=0,
\end{equation}
needs to be satisfied for the fields; also see \cite{KR2}.
Using (\ref{Feldnull}), we get $\Delta B_1=-4\pi\curl j_0$ and therefore define
\begin{equation}\label{DefBe}
   B_1(x, t)=\int |x-y|^{-1}\curl j_0(y, t)\,dy
   =\int |z|^{-2}\bar{z}\times j_0(x+z, t)\,dz.
\end{equation}
Regarding the density $f_1$, we obtain the linear Vlasov equation
\[ \partial_t f_1+v\cdot\nabla_x f_1+E_1\cdot\nabla_v f_0+E_0\cdot\nabla_v f_1=0. \]
Hence if we suppose that $f_1(x, v, 0)=0$, then we can set
\[ f_1=0\quad\mbox{and}\quad E_1=0 \]
consistently.

The field equations at the order $c^{-2}$ are
\[ \curl E_2=-\partial_t B_1,\quad\div E_2=4\pi\rho_2,
   \quad\curl B_2=\partial_t E_1+4\pi j_1=0,\quad\div B_2=0. \]
Therefore we can define
\[ B_2=0. \]
Calculating the equation for the density $f_2$ and taking into account (\ref{Feldeins}),
we arrive at the following inhomogeneous linearized Vlasov-Poisson system,
for which we choose homogeneous initial data:
\begin{equation}\label{LVP}
   \left\{\begin{array}{c}
   \partial_t f_2+v\cdot\nabla_x f_2-\frac{1}{2} v^2\,v\cdot\nabla_x f_0
   +E_0\cdot\nabla_v f_2+(E_2+v\times B_1)\cdot\nabla_v f_0=0, \\[1ex]
   \Delta E_2=\partial_t^2 E_0+4\pi(\nabla\rho_2+\partial_t j_0), \\[1ex]
   f_2(x, v, 0)=0.\end{array}\right.\tag{LVP}
\end{equation}
At this point we need to discuss the solvability of the Poisson equation for $E_2$.
Restricting our attention to initial data $f^\circ$ for (\ref{RVMC})
with compact support, it will turn out below that both $\rho_2$ and $j_0$
have compact support and thus lead to unproblematic sources.
The first term $\partial_t^2 E_0$ has to be examined more closely.
Since $\rho_0(\cdot, t)$ has compact support for all $t$,
see (\ref{SchrankeGeschwVP}) below, we can calculate the iterated Poisson integrals
\begin{eqnarray}
   \Delta^{-1}(\partial_t^2 E_0)(x, t)
   & = & -\frac{1}{4\pi}\int\frac{dy}{|x-y|}\,\partial_t^2 E_0(y, t)
   \nonumber \\ & = & \frac{1}{4\pi}\int\frac{dw}{|w|}
   \,\int |z|^{-2}\bar{z}\,\partial_t^2\rho_0(x+w+z, t)\,dz
   \nonumber \\ & = & \frac{1}{4\pi}\int dy\,\partial_t^2\rho_0(y, t)
   \int du\,|y-x-u|^{-1}|u|^{-3}u
   \nonumber \\ & = & \frac{1}{2}\int\frac{dy}{|y-x|}(y-x)\,\partial_t^2\rho_0(y, t)
   =\frac{1}{2}\int\bar{z}\,\partial_t^2\rho_0(x+z, t)\,dz
   \label{unpr2} \\ & = & \frac{1}{2}\int\int |z|^{-1}(v-(\bar{z}\cdot v)\bar{z})
   \,\partial_t f_0(x+z, v, t)\,dz\,dv, \nonumber
\end{eqnarray}
where we used (\ref{VP}), eq.~(\ref{formelv}) from the appendix,
and $\partial_t\rho_0+\div j_0=0$ in conjunction with an integration by parts,
the continuity equation itself being a direct consequence of (\ref{VP}).
In view of (\ref{unpr2}) and (\ref{LVP}) we thus define
\begin{eqnarray}\label{DefEz}
   E_2(x, t) & = & \frac{1}{2}\int\bar{z}\,\partial_t^2\rho_0(x+z, t)\,dz
   -\int |z|^{-1}\partial_t j_0(x+z, t)\,dz
   -\int |z|^{-2}\bar{z}\,\rho_2(x+z, t)\,dz. \qquad
\end{eqnarray}
By (\ref{DefEz}), (\ref{VP}), and a further integration by parts,
we obtain the alternative expression
\begin{eqnarray}\label{ADarEz}
    E_2(x, t) & = & \frac{1}{2}\int\int|z|^{-2}\bar{z}\,
    (3(\bar{z}\cdot v)^2-v^2)\,f_0(x+z, v, t)\,dz\,dv \nonumber \\
    & & -\frac{1}{2}\int |z|^{-1}(1+\bar{z}\otimes\bar{z})
    \,(E_0\rho_0)(x+z, t)\,dz-\int |z|^{-2}\bar{z}\,\rho_2(x+z, t)\,dz.
\end{eqnarray}
The first aim of this paper is to show that
\begin{eqnarray}
   f^D & := & f_0+c^{-2}f_2, \nonumber \\
   E^D & := & E_0+c^{-2}E_2, \label{DefDarwin} \\
   B^D & := & c^{-1}B_1, \nonumber
\end{eqnarray}
yields a higher-order pointwise approximation of (\ref{RVMC})
than the Vlasov-Poisson system; we call (\ref{DefDarwin}) the Darwin approximation.
It is clear that for achieving this improved approximation property
also the initial data of (\ref{RVMC}) have to be matched appropriately
by the data for the Darwin system. For a prescribed initial density $f^\circ$,
we are able to calculate $(f_0, E_0)$, $B_1$, and $(f_2, E_2)$
according to what has been outlined above. We then consider (\ref{RVMC})
with initial data
\begin{equation}\label{IC}\tag{IC}
   \left\{\begin{array}{lcl}f(x, v, 0) & = & f^\circ(x, v), \\
   E(x, 0) & = & E^\circ(x):=E_0(x, 0)+c^{-2}E_2(x, 0), \\
   B(x, 0) & = & B^\circ(x):=c^{-1}B_1(x, 0). \end{array}\right.
\end{equation}
Before we formulate our main theorem let us recall that solutions of (\ref{RVMC})
with initial data (\ref{IC}) exist at least on some time interval $[0, T]$
which is independent of $c\geq 1$; see \cite[Thm.~1]{schaeffer:86},
and cf.~Proposition \ref{Schaeffer-prop} below for a more precise statement.
This time interval $[0, T]$ is fixed throughout the paper.

\begin{theorem}\label{Hauptsatz}
Assume that $f^\circ\in C^\infty(\R^3\times\R^3)$ is nonnegative
and has compact support. From $f^\circ$ calculate $(f_0, E_0)$, $B_1$,
and $(f_2, E_2)$, and then define initial data for (\ref{RVMC})
by (\ref{IC}). Let $(f, E, B)$ denote the solution of (\ref{RVMC})
with initial data (\ref{IC}) and let $(f^D, E^D, B^D)$ be defined
as in (\ref{DefDarwin}). Then there exists a constant $M>0$,
and also for every $R>0$ there is $M_R>0$, such that
\begin{eqnarray}
   |f(x, v, t)-f^D(x, v, t)| & \leq & Mc^{-3}\quad\hspace{0.55em} (x\in\R^3), 
   \nonumber \\
   |E(x, t)-E^D(x, t)| & \leq & M_R\,c^{-3}\quad (|x|\le R), \label{diff-esti} \\
   |B(x, t)-B^D(x, t)| & \leq & Mc^{-3}\quad\hspace{0.65em} (x\in\R^3), \nonumber
\end{eqnarray}
for all $v\in\R^3$, $t\in [0, T]$, and $c\geq 1$.
\end{theorem}

The constants $M$ and $M_R$ are independent of $c\geq 1$,
but do depend on the initial data. Note that if (\ref{RVMC})
is compared to the Vlasov-Poisson system (\ref{VP}) only,
one obtains the estimate $|f(x, v, t)-f_0(x, v, t)|
+|E(x, t)-E_0(x, t)|+|B(x, t)|\le Mc^{-1}$; see \cite[Thm.~2B]{schaeffer:86}.

Approximate models have the big advantage that, since by now the Vlasov-Poisson system
is well understood, the existence of $(f_0, E_0)$, and here also of $B_1$
and $(f_2, E_2)$, does no longer pose serious problems; note that in (\ref{LVP})
the equation for $f_2$ is linear. Therefore one can hope to get more information
on (\ref{RVMC}) by studying the approximate equations. As a drawback
of the above hierarchy, one has to deal with two densities $f_0$, $f_2$
and two electric fields $E_0$, $E_2$ to define $f^D$ and $E^D$. Therefore 
it is natural to look for a model which can be written down using only one density 
and one field. It turns out that the appropriate (Hamiltonian) system is
\begin{equation}\label{DaVlaMax}
   \left\{\begin{array}{l}\partial_t f+(1-\frac{1}{2}\,c^{-2}v^2)v\cdot\nabla_x f
   +(E+c^{-1}v\times B)\cdot\nabla_v f=0, \\[1ex]
   c\curl E=-\partial_t B,\quad\div E=4\pi\rho, \\
   c\,\Delta B=-4\pi\curl j, \\[1ex]
   \rho=\int f\,dv,\quad j=\int (1-\frac{1}{2}\,c^{-2}v^2)v\,f\,dv,
   \end{array}\right.\tag{DVMc}
\end{equation}
which we call the Darwin-Vlasov-Maxwell system. We note that $(f^D, E^D, B^D)$
solves (\ref{DaVlaMax}) up to an error of the order $c^{-3}$.

\begin{theorem}\label{DVM-thm} Assume that $f^\circ\in C^\infty(\R^3\times\R^3)$
is nonnegative and has compact support. Then there exist $c^\ast\ge 1$ and $T^\ast>0$
such that the following holds for $c\geq c^\ast$.
\begin{itemize}
\item[(a)] If there is a local solution of (\ref{DaVlaMax}), then the initial data
$E^\circ$ and $B^\circ$ of (\ref{DaVlaMax}) at $t=0$ are uniquely determined
by the initial density $f^\circ$.
\item[(b)] The system (\ref{DaVlaMax}) has a unique $C^2$-solution
$(f^\ast, E^\ast, B^\ast)$ on $[0, T^\ast]$ attaining that initial data
$(f^\circ, E^\circ, B^\circ)$ at $t=0$. This solution conserves the energy 
\[ {\cal H}=\int\int\Big(\frac{1}{2}\,v^2-\frac{1}{8}\,c^{-2}v^4\Big)f^\ast\,dx\,dv
   +\frac{1}{8\pi}\int\Big(|\nabla\phi^\ast|^2+|\nabla\wedge A^\ast|^2\Big)\,dx, \] 
where the potentials $\phi^\ast$ and $A^\ast$ are chosen in such a way that 
$B^\ast=\nabla\wedge A^\ast$, $\nabla\cdot A^\ast=0$, 
and $-\nabla\phi^\ast=E^\ast+c^{-1}\partial_t A^\ast$. 
\item[(c)] Let $(f, E, B)$ denote the solution of (\ref{RVMC})
with initial data $(f^\circ, E^\circ, B^\circ)$. Then there exists
a constant $M>0$, and also for every $R>0$ there is $M_R>0$, such that
\begin{eqnarray*}
   |f(x, v, t)-f^\ast(x, v, t)| & \leq & Mc^{-3}\quad\hspace{0.55em} (x\in\R^3), \\
   |E(x, t)-E^\ast(x, t)| & \leq & M_R\,c^{-3}\quad (|x|\le R), \\
   |B(x, t)-B^\ast(x, t)| & \leq & Mc^{-3}\quad\hspace{0.6em} (x\in\R^3),
\end{eqnarray*}
for all $v\in\R^3$, $t\in [0, \min\{T, T^\ast\}]$, and $c\geq c^\ast$.
\end{itemize}
\end{theorem}

Instead of performing the limit $c\to\infty$ in (\ref{RVMC}) it is possible
to reformulate Theorem \ref{Hauptsatz} in terms of a suitable dimensionless parameter.
Taking this viewpoint means that we consider (\ref{RVMC}) at a fixed $c$ (say $c=1$)
by rescaling a prescribed nonnegative initial density $f^\circ$,
for which we suppose that $f^\circ\in C^\infty(\R^3\times\R^3)$
has compact support. To be more precise, let
\[ \bar{v}=\int\int\hat{v} f^\circ(x, v)\,dx\,dv, \]
where $\hat{v}$ is taken for $c=\eps^{-1/2}$; cf.~(\ref{hatv-def}).
Then $\bar{v}$ is viewed as an average velocity of the system.
Now we introduce
\[ f^{\eps, \circ}(x, v)=\eps^{3/2} f^\circ(\eps x, \eps^{-1/2} v) \]
and consider $f^{\eps, \circ}$ for $c=1$. It follows that
\[ \bar{v}^\eps=\int\int\hat{v} f^{\eps, \circ}(x, v)\,dx\,dv
   =\sqrt{\eps}\int\int\hat{w} f^\circ(y, w)\,dy\,dw=\sqrt{\eps}\,\bar{v}, \]
i.e., the system with initial distribution function $f^{\eps, \circ}$
has small velocities compared to the system associated to $f^\circ$.
Starting from $f^\circ$, we next determine $(f_0, E_0)$, $B_1$,
and $(f_2, E_2)$, and then the initial data for (\ref{RVMC}) via (\ref{IC})
with $c=\eps^{-1/2}$, as in Theorem \ref{Hauptsatz}. Next we note that $(f, E, B)$
is a solution of (\ref{RVMC}) with $c=\eps^{-1/2}$ if and only if
\begin{eqnarray*}
   f^\eps(x, v, t) & = & \eps^{3/2} f(\eps x, \eps^{-1/2} v, \eps^{3/2} t),
   \\ E^\eps(x, t) & = & \eps^2 E(\eps x, \eps^{3/2} t),
   \\ B^\eps(x, t) & = & \eps^2 B(\eps x, \eps^{3/2} t),
\end{eqnarray*}
is a solution of (\ref{RVMC}) with $c=1$. We further introduce
\begin{eqnarray*}
   f_0^\eps(x, v, t) & = & \eps^{3/2} f_0(\eps x, \eps^{-1/2} v, \eps^{3/2} t), \\
   E_0^\eps(x, t) & = & \eps^2 E_0(\eps x, \eps^{3/2} t), \\
   B_1^\eps(x, t) & = & \eps^{5/2} B_1(\eps x, \eps^{3/2} t), \\
   f_2^\eps(x, v, t) & = & \eps^{5/2} f_2(\eps x, \eps^{-1/2} v, \eps^{3/2} t), \\
   E_2^\eps(x, t) & = & \eps^3 E_2(\eps x, \eps^{3/2} t), \\
   \rho_0^\eps(x, t) & = & \int f_0^\eps(x, v, t)\,dv
   =\eps^3\rho_0(\eps x, \eps^{3/2} t), \\
   j_0^\eps(x, t) & = & \int v f_0^\eps(x, v, t)\,dv
   =\eps^{7/2} j_0(\eps x, \eps^{3/2} t), \\
   \rho_2^\eps(x, t) & = & \int f_2^\eps(x, v, t)\,dv
   =\eps^4\rho_2(\eps x, \eps^{3/2} t).
\end{eqnarray*}
Straightforward calculations then confirm the following statements:
\begin{itemize}
\item[(a)] $(f_0, E_0)$ is a solution to (\ref{VP}) with initial data $f^\circ$
if and only if $(f_0^\eps, E_0^\eps)$ is a solution to (\ref{VP})
with initial data $f^{\eps, \circ}$,
\item[(b)] $B_1$ solves $\Delta B_1=-4\pi\curl j_0$ if and only if $B_1^\eps$
solves $\Delta B_1^\eps=-4\pi\curl j_0^\eps$,
\item[(c)] $(f_2, E_2)$ is a solution to (\ref{LVP}) if and only
if $(f_2^\eps, E_2^\eps)$ is a solution to
\[ \left\{\begin{array}{c}
   \partial_t f_2^\eps+v\cdot\nabla_x f_2^\eps
   -\frac{1}{2} v^2\,v\cdot\nabla_x f_0^\eps
   +E_0^\eps\cdot\nabla_v f_2^\eps+(E_2^\eps+v\times B_1^\eps)
   \cdot\nabla_v f_0^\eps=0, \\[1ex]
   \Delta E_2^\eps=\partial_t^2 E_0^\eps+4\pi(\nabla\rho_2^\eps
   +\partial_t j_0^\eps), \\[1ex]
   f_2^\eps(x, v, 0)=0.\end{array}\right. \]
\end{itemize}
Therefore Theorem \ref{Hauptsatz} may be reformulated in a way which parallels
\cite[Thm.~2.2]{KS1}, where the case of individual particles is considered
which are far apart (of order ${\cal O}(\eps^{-1})$) and have small velocities
(of order ${\cal O}(\sqrt{\eps})$) initially. Note that in this result
the Lorentz force is determined up to an error of order ${\cal O}(\eps^{7/2})$,
and the dynamics of the full and the effective system can be compared over long times
of order ${\cal O}(\eps^{-3/2})$; see \cite[p.~448]{KS1}.

\begin{theorem}\label{eps-thm}
Assume that $f^\circ\in C^\infty(\R^3\times\R^3)$ is nonnegative
and has compact support. From $f^\circ$ calculate $(f_0, E_0)$, $B_1$,
and $(f_2, E_2)$, and then define initial data for (\ref{RVMC})
by (\ref{IC}) with $c=\eps^{-1/2}$. Let $(f, E, B)$ denote the solution of (\ref{RVMC})
on $[0, T]$ for $c=\eps^{-1/2}$ with initial data (\ref{IC}). Moreover,
let $f^{\eps, \circ}$, $f^\eps$, $E^\eps$, $B^\eps$, $f_0^\eps$, $E_0^\eps$,
$B_1^\eps$, $f_2^\eps$, $E_2^\eps$, $\rho_0^\eps$, $j_0^\eps$, and $\rho_2^\eps$
be defined as above. Then $(f^\eps, E^\eps, B^\eps)$ is a solution of (\ref{RVMC})
on $[0, \eps^{-3/2}T]$ for $c=1$ with initial data
\[ (f^\eps, E^\eps, B^\eps)(x, v, 0)=(f^{\eps, \circ}(x, v),
   E_0^\eps(x, 0)+E_2^\eps(x, 0), B_1^\eps(x, 0)). \]
In addition, there exists a constant $M>0$, and also for every $R>0$
there is $M_R>0$, such that
\begin{eqnarray*}
   |f^\eps(x, v, t)-f_0^\eps(x, v, t)-f_2^\eps(x, v, t)| & \leq & M\eps^3
   \quad\hspace{1.3em} (x\in\R^3),\\
   |E^\eps(x, t)-E_0^\eps(x, t)-E_2^\eps(x, t)| & \leq & M_R\,\eps^{7/2}
   \quad (|x|\le \eps^{-1}R), \\
   |B^\eps(x, t)-B_1^\eps(x, t)| & \leq & M\eps^{7/2}
   \quad\hspace{0.6em} (x\in\R^3),
\end{eqnarray*}
for all $v\in\R^3$, $t\in [0, \eps^{-3/2}T]$, and $\eps\leq 1$.
The constants are independent of $\eps$.
\end{theorem}
By definition of the rescaled fields, these fields are slowly
varying in their space and time variables, which means that we are considering
an adiabatic limit. It is clear that also Theorem \ref{DVM-thm} could be restated
in an analogous $\eps$-dependent version.

The paper is organized as follows. Some facts concerning (\ref{VP}),
(\ref{LVP}), and (\ref{RVMC}) are collected in Section \ref{eff-syst}.
The proof of Theorem \ref{Hauptsatz} is elaborated in Section \ref{HS-bew},
whereas Section \ref{DVM-bew} contains the proof of Theorem \ref{DVM-thm}.
For the proofs we will mostly rely on suitable representation formulas
for the fields (refined versions of those used in \cite{glstr,schaeffer:86}),
which are derived in the appendix, Section \ref{append}.
\smallskip

\noindent
{\bf Notation:} $B(0, R)$ denotes the closed ball in $\R^3$
with center at $x=0$ or $v=0$ and radius $R>0$.
The usual $L^\infty$-norm of a function $\varphi=\varphi(x)$ over $x\in\R^3$
is written as ${\|\varphi\|}_x$, and if $\varphi=\varphi(x, v)$,
we modify this to ${\|\varphi\|}_{x, v}$.
For $m\in\N$ the $W^{m, \infty}$-norms are denoted by ${\|\varphi\|}_{m, x}$, etc.
If $T>0$ is fixed, then we write
\[ g(x, v, t, c)={\cal O}_{cpt}(c^{-m}), \]
if for all $R>0$ there is a constant $M=M_R>0$ such that
\begin{equation}\label{GOForm}
   |g(x, v, t, c)|\leq Mc^{-m}
\end{equation}
for $|x|\le R$, $v\in\R^3$, $t\in [0, T]$, and $c\geq 1$.
Similarly, we write
\[ g(x, v, t, c)={\cal O}(c^{-m}), \]
if there is a constant $M>0$ such that (\ref{GOForm}) holds for all
$x, v\in\R^3$, $t\in [0, T]$, and $c\geq 1$. In general, generic constants
are denoted by $M$.


\setcounter{equation}{0}

\section{Some properties of (\ref{VP}), (\ref{LVP}), and (\ref{RVMC})}
\label{eff-syst}

There is a vast literature on (\ref{VP}), see e.g.~\cite[Sect.~4]{glassey:96}
or \cite{rein} and the references therein. For our purposes we collect
a few well known facts about classical solutions of (\ref{VP}).

\begin{proposition} Assume that $f^\circ\in C^\infty(\R^3\times\R^3)$
is nonnegative and has compact support. Then there exists a unique global
$C^1$-solution $(f_0, E_0)$ of (\ref{VP}), and there are nondecreasing
continuous functions $P_{V\!P}, K_{V\!P}: [0, \infty[\to\R$ such that
\begin{gather}
   {\|f_0(t)\|}_{x, v}\leq {\|f^\circ\|}_{x, v}, \nonumber \\
   {\rm supp}\,f_0(\cdot, \cdot, t)
   \subset B(0, P_{V\!P}(t))\times B(0, P_{V\!P}(t)),
   \label{SchrankeGeschwVP} \\
   {\|f_0(t)\|}_{1, x, v}+{\|E_0(t)\|}_{1, x}\leq K_{V\!P}(t), \nonumber
\end{gather}
for $t\in [0, \infty[$.
\end{proposition}

This result was first established by Pfaffelmoser
\cite{pfaffelmoser:92}, and simplified versions of the proof
were obtained by Schaeffer \cite{schaeffer:91} and Horst \cite{horst:93};
a proof along different lines is due to Lions and Perthame \cite{lions/perthame:91}.

For our approximation scheme we also need bounds on higher derivatives
of the solution. This point was elaborated in \cite{lindner:91},
where it was shown that if $f^\circ\in C^k(\R^3\times\R^3)$,
then $(f_0, E_0)$ posses continuous partial derivatives w.r.t.~$x$ and $v$
up to order $k$. The existence of continuous time-derivatives then
follows from the Vlasov equation. Thus $(f_0, E_0)$ are $C^\infty$,
if $f^\circ$ is $C^\infty$, and by a redefinition of $K_{V\!P}$ we can assume that
\begin{equation}\label{SchrankeFeldVP}
   {\|f_0(t)\|}_{3, x, v}\leq K_{V\!P}(t),\quad t\in [0, \infty[.
\end{equation}

The existence of a unique $C^1$-solution $(f_2, E_2)$ of (\ref{LVP})
follows by a contraction argument, but we omit the details. Furthermore
it can be shown that there are nondecreasing continuous functions
$P_{LV\!P}, K_{LV\!P}: [0, \infty[\to\R$ such that
\begin{gather}
   \label{TrLVP}
   {\rm supp}\,f_2(\cdot, \cdot, t)
   \subset B(0, P_{LV\!P}(t))\times B(0, P_{LV\!P}(t)), \\
   \label{SchrankeLVP}
   {\|f_2(t)\|}_{1, x, v}+{\|E_2(t)\|}_{1, x}\leq K_{LV\!P}(t),
\end{gather}
for $t\in [0, \infty[$.

Concerning solutions of (\ref{RVMC}), we have from \cite[Thm.~1]{schaeffer:86}
the following

\begin{proposition}\label{Schaeffer-prop}
Assume that $f^\circ\in C^\infty(\R^3\times\R^3)$ is nonnegative
and has compact support. If $E^\circ$ and $B^\circ$ are defined by (\ref{IC}),
then there exits $T>0$ (independent of $c$) such that for all $c\geq 1$
the system (\ref{RVMC}) with initial data (\ref{IC}) has a unique $C^1$-solution
$(f, E, B)$ on the time interval $[0, T]$. In addition, there are
nondecreasing continuous functions (independent of $c$)
$P_{V\!M}, K_{V\!M}: [0, T]\to\R$ such that
\begin{gather}
   f(x, v, t) = 0\quad\mbox{if}\quad |v|\geq P_{V\!M}(t),
   \label{SchrankeSupport} \\
   |E(x, t)|+|B(x, t)|\leq K_{V\!M}(t),
   \label{SchrankeFelder}
\end{gather}
for all $x\in\R^3$, $t\in [0, T]$, and $c\geq 1$.
\end{proposition}

In fact $E^\circ$ and $B^\circ$ do not depend on $c$ in \cite[Thm.~1]{schaeffer:86},
but an inspection of the proof shows that the assertions remain valid
for initial fields defined by (\ref{IC}).


\setcounter{equation}{0}

\section{Proof of Theorem \ref{Hauptsatz}}
\label{HS-bew}

In Section \ref{repapp-sect} below we will show that the approximate electric field
$E^D$ from (\ref{DefDarwin}) admits the following representation:

\begin{equation}\label{DarstellungED}
   E^D=E^D_{{\rm ext}}+E^D_{{\rm int}}+E^D_{{\rm bd}}+{\cal O}_{cpt}(c^{-3}),
\end{equation}
with
\begin{eqnarray}
   E^D_{{\rm ext}}(x, t) & = & -\int_{|z|>ct} |z|^{-2}\bar{z}\,(\rho_0+c^{-2}\rho_2)(x+z, t)\,dz
   \nonumber \\
   & & -c^{-2}\int_{|z|>ct} |z|^{-1}\partial_t j_0(x+z, t)\,dz
   +\frac{1}{2}\,c^{-2}\int_{|z|>ct}\bar{z}\,\partial_t^2\rho_0(x+z, t)\,dz, \label{ED-ext} \\
   E^D_{{\rm int}}(x, t) & = & -\int_{|z|\leq ct}|z|^{-2}\bar{z}
   \,(\rho_0+c^{-2}\rho_2)(x+z, \hat{t}(z))\,dz \nonumber \\
   & & -c^{-1}\int_{|z|\leq ct}\int |z|^{-2}(v-2(\bar{z}\cdot v)\bar{z})
   \,f_0(x+z, v, \hat{t}(z))\,dv\,dz \nonumber \\
   & & +c^{-2}\int_{|z|\leq ct}\int |z|^{-2}(2(\bar{z}\cdot v)v
   +v^2\bar{z}-3\bar{z}(\bar{z}\cdot v)^2)\,f_0(x+z, v, \hat{t}(z))\,dv\,dz \nonumber \\
   & & +c^{-2}\int_{|z|\leq ct}|z|^{-1}(\bar{z}\otimes\bar{z}-1)E_0
   \rho_0(x+z, \hat{t}(z))\,dz, \label{ED-int} \\
   E^D_{{\rm bd}}(x, t) & = & c^{-1}(ct)^{-1}\int_{|z|=ct}\int (\bar{z}\cdot v)\bar{z}
   \,f^\circ(x+z, v)\,dv\,ds(z) \nonumber \\
   & & +c^{-2}(ct)^{-1}\int_{|z|=ct}\int((\bar{z}\cdot v)v
   -(\bar{z}\cdot v)^2\bar{z})\,f^\circ(x+z, v)\,dv\,ds(z), \nonumber
\end{eqnarray}
where the subscripts `ext', `int', and `bd' refer to the exterior, interior,
and boundary integration in $z$. We also recall that $\bar{z}=|z|^{-1}z$
and $\hat{t}(z)=t-c^{-1}|z|$.
On the other hand, according to Section \ref{max-repres} below we have
\begin{equation}\label{DarstellungE}
   E=E_{{\rm ext}}+E_{{\rm int}}+E_{{\rm bd}}+{\cal O}(c^{-3}),
\end{equation}
with
\begin{eqnarray}
   E_{{\rm ext}}(x, t) & = & -\int_{|z|>ct} |z|^{-2}\bar{z}
   \,\Big(\rho_0+t\partial_t\rho_0+\frac{1}{2}\,t^2\partial_t^2\rho_0\Big)(x+z, 0)\,dz
   \nonumber \\ & & +\frac{1}{2}\,c^{-2}\int_{|z|>ct}\bar{z}\,\partial_t^2\rho_0(x+z, 0)\,dz
   -c^{-2}\int_{|z|>ct}|z|^{-1}\partial_t j_0(x+z, 0)\,dz, \label{E-ext} \\
   E_{{\rm int}}(x, t) & = & -\int_{|z|\leq ct} |z|^{-2}\bar{z}\rho(x+z, \hat{t}(z))\,dz
   \nonumber \\
   & & +c^{-1}\int_{|z|\leq ct} |z|^{-2}\int (2(\bar{z}\cdot v)\bar{z}-v)
   \,f(x+z, v, \hat{t}(z))\,dv\,dz \nonumber \\
   & & +c^{-2}\int_{|z|\leq ct} |z|^{-2}\int (v^2\bar{z}+2(\bar{z}\cdot v)v
   -3(\bar{z}\cdot v)^2\bar{z})\,f(x+z, v, \hat{t}(z))\,dv\,dz  \nonumber \\
   & & +c^{-2}\int_{|z|\leq ct} |z|^{-1}\int (\bar{z}\otimes\bar{z}-1)
   (Ef)(x+z, v, \hat{t}(z))\,dv\,dz, \label{E-int} \\
   E_{{\rm bd}}(x, t) & = & E^D_{{\rm bd}}(x, t). \nonumber
\end{eqnarray}

In order to verify (\ref{diff-esti}), we start by comparing the exterior fields.
Let $x\in B(0, R)$ with $R>0$ be fixed. Then we obtain from (\ref{E-ext}) and (\ref{ED-ext}),
due to $|\bar{z}|=1$, and taking into account
\[ \rho_2(x, 0)=\int f_2(x, v, 0)\,dv=0 \]
by (\ref{LVP}), as well as (\ref{SchrankeGeschwVP}), (\ref{SchrankeFeldVP}),
(\ref{TrLVP}), and (\ref{SchrankeLVP}),
\begin{eqnarray}\label{ext-diff}
   \lefteqn{|E_{{\rm ext}}(x, t)-E^D_{{\rm ext}}(x, t)|} \nonumber
   \\ & \le & \int_{|z|>ct} |z|^{-2}\Big|\rho_0(x+z, t)-\rho_0(x+z, 0)-t\partial_t\rho_0(x+z, 0)
   -\frac{1}{2}\,t^2\partial_t^2\rho_0(x+z, 0)\Big|\,dz \nonumber
   \\ & & +c^{-2}\int_{|z|>ct} |z|^{-2}|\rho_2(x+z, t)-\rho_2(x+z, 0)|\,dz \nonumber
   \\ & & +c^{-2}\int_{|z|>ct}|z|^{-1}\int |v|\,|\partial_t f_0(x+z, v, 0)
   -\partial_t f_0(x+z, v, t)|\,dv\,dz \nonumber
   \\ & & +\frac{1}{2}\,c^{-2}\int_{|z|>ct}\int |\partial_t^2 f_0(x+z, v, t)
   -\partial_t^2 f_0(x+z, v, 0)|\,dv\,dz \nonumber
   \\ & \le & M\int_{|z|>ct} |z|^{-2}\bigg(\int_0^t(t-s)^2 P_{V\!P}(s)^3 K_{V\!P}(s)
   {\bf 1}_{B(0, P_{V\!P}(s))}(x+z)\,ds\bigg)\,dz \nonumber
   \\ & & +Mc^{-2}\int_{|z|>ct} |z|^{-2}\bigg(\int_0^t P_{LV\!P}(s)^3 K_{LV\!P}(s)
   {\bf 1}_{B(0, P_{LV\!P}(s))}(x+z)\,ds\bigg)\,dz \nonumber
   \\ & & +Mc^{-2}\int_{|z|>ct}|z|^{-1}\bigg(\int_0^t P_{V\!P}(s)^4 K_{V\!P}(s)
   {\bf 1}_{B(0, P_{V\!P}(s))}(x+z)\,ds\bigg)\,dz \nonumber
   \\ & & +Mc^{-2}\int_{|z|>ct}\bigg(\int_0^t P_{V\!P}(s)^3 K_{V\!P}(s)
   {\bf 1}_{B(0, P_{V\!P}(s))}(x+z)\,ds\bigg)\,dz \nonumber
   \\ & \le & Mt^3\int_{|z|>ct} |z|^{-2}{\bf 1}_{B(0, R+M_0)}(z)\,dz
   +Mt\,c^{-2}\int_{|z|>ct} |z|^{-1}(|z|^{-1}+1+|z|){\bf 1}_{B(0, R+M_0)}(z)\,dz \nonumber
   \\ & \le & M_R\,c^{-3};
\end{eqnarray}
note that here we have used
\begin{equation}\label{M0-def}
   M_0=\max_{s\in [0, T]}\Big(P_{V\!P}(s)+K_{V\!P}(s)+P_{LV\!P}(s)
   +K_{LV\!P}(s)\Big)<\infty,
\end{equation}
and for instance
\[ t^3\int_{|z|>ct} |z|^{-2}{\bf 1}_{B(0, R+M_0)}(z)\,dz
   \le (ct)^{-3}t^3\int_{|z|\le R+M_0} |z|\,dz\le M_R\,c^{-3}. \]

To bound $|E_{{\rm int}}(x, t)-E^D_{{\rm int}}(x, t)|$, we first recall from
\cite[Thm.~2B]{schaeffer:86} that
\begin{equation}\label{ma6}
   |E(x, t)-E_0(x, t)|={\cal O}(c^{-1}).
\end{equation}
Actually the initial conditions in \cite{schaeffer:86} are different,
but we only added terms of order $c^{-2}$, so that an inspection of the proof
in \cite{schaeffer:86} leads to (\ref{ma6}). Next we define
\[ H(t)=\sup\,\{|f(x, v, s)-f^D(x, v, s)|: x\in\R^3, v\in\R^3, s\in [0, t]\}, \]
as well as
\[ M_1=\max_{s\in [0, T]}\Big(P_{V\!M}(s)+P_{V\!P}(s)+P_{LV\!P}(s)\Big)<\infty. \]
Then $f(x, v, s)=f_0(x, v, s)=f_2(x, v, s)=0$ for $x\in\R^3$, $|v|\ge M_1$, and $s\in [0, T]$.
Also if $R_0>0$ is chosen such that $f^\circ(x, v)=0$ for $|x|\ge R_0$, introducing the constant
\[ M_2=R_0+TM_1+\max_{s\in [0, T]}\Big(P_{V\!P}(s)+P_{LV\!P}(s)\Big)<\infty  \]
it follows that $f(x, v, s)=f_0(x, v, s)=f_2(x, v, s)=0$
for $|x|\ge M_2$, $v\in\R^3$, and $s\in [0, T]$. Let $x\in B(0, R)$ with $R>0$ be fixed.
From (\ref{E-int}), (\ref{ED-int}), (\ref{M0-def}), (\ref{ma6}), (\ref{IC}),
and $0\le\hat{t}(z)\le t$ for $|z|\le ct$ we obtain
\begin{eqnarray}\label{int-diff}
   \lefteqn{|E_{{\rm int}}(x, t)-E^D_{{\rm int}}(x, t)|} \nonumber
   \\ & \le & \int_{|z|\leq ct} |z|^{-2}\,\bigg|\int (f-f_0-c^{-2}f_2)
   (x+z, v, \hat{t}(z))\,dv\bigg|\,dz \nonumber \\
   & & +c^{-1}\int_{|z|\leq ct} |z|^{-2}\bigg|\int (2(\bar{z}\cdot v)\bar{z}-v)
   \,(f-f_0-c^{-2}f_2)(x+z, v, \hat{t}(z))\,dv\bigg|\,dz \nonumber \\
   & & +c^{-3}\int_{|z|\leq ct} |z|^{-2}\bigg|\int (2(\bar{z}\cdot v)\bar{z}-v)
   \,f_2(x+z, v, \hat{t}(z))\,dv\bigg|\,dz \nonumber \\
   & & +c^{-2}\int_{|z|\leq ct} |z|^{-2}\bigg|\int (v^2\bar{z}+2(\bar{z}\cdot v)v
   -3(\bar{z}\cdot v)^2\bar{z})\,(f-f_0-c^{-2}f_2)(x+z, v, \hat{t}(z))\,dv\bigg|\,dz
   \nonumber \\
   & & +c^{-4}\int_{|z|\leq ct} |z|^{-2}\bigg|\int (v^2\bar{z}+2(\bar{z}\cdot v)v
   -3(\bar{z}\cdot v)^2\bar{z})\,f_2(x+z, v, \hat{t}(z))\,dv\bigg|\,dz \nonumber \\
   & & +c^{-2}\int_{|z|\leq ct} |z|^{-1}\bigg|\int (1-\bar{z}\otimes\bar{z})
   ([E-E_0]f)(x+z, v, \hat{t}(z))\,dv\bigg|\,dz
   \nonumber \\ & \le & M (M_1^3+M_1^4) H(t)\int_{|z|\leq ct} |z|^{-2}
   \,{\bf 1}_{B(0, M_2)}(x+z)\,dz\nonumber \\
   & & +M M_1^4 M_0\,c^{-3}\int_{|z|\leq ct} |z|^{-2}\,{\bf 1}_{B(0, M_2)}(x+z)\,dz
   \nonumber \\
   & & +M M_1^5 H(t)\,c^{-2}\int_{|z|\leq ct} |z|^{-2}\,{\bf 1}_{B(0, M_2)}(x+z)\,dz
   \nonumber \\
   & & +M M_1^5 M_0\,c^{-4}\int_{|z|\leq ct} |z|^{-2}\,{\bf 1}_{B(0, M_2)}(x+z)\,dz
   \nonumber \\
   & & +M M_1^3 {\|f^\circ\|}_{x, v}\,c^{-3}\int_{|z|\leq ct} |z|^{-1}
   \,{\bf 1}_{B(0, M_2)}(x+z)\,dz \nonumber \\
   & \le & M_R(c^{-3}+H(t)),
\end{eqnarray}
since for instance
\[ \int_{|z|\leq ct} |z|^{-2}\,{\bf 1}_{B(0, M_2)}(x+z)\,dz
   \le\int_{|z|\leq R+M_2} |z|^{-2}\,dz\le M_R. \]
Recalling that the $E_{{\rm bd}}(x, t)=E^D_{{\rm bd}}(x, t)$,
we can summarize (\ref{DarstellungE}), (\ref{DarstellungED}), (\ref{ext-diff}),
and (\ref{int-diff}) as
\begin{equation}\label{abschE}
   |E(x, t)-E^D(x, t)|\le M_R(c^{-3}+H(t)),
\end{equation}
for $|x|\le R$ and $t\in [0, T]$. Formulas (\ref{RepBD}), (\ref{BFeld}),
(\ref{BD-form}), (\ref{BDT-expa}), (\ref{BT-expa}), and (\ref{BS-expa}),
and an analogous (actually more simple) calculation also leads to
\begin{equation}\label{abschB}
   |B(x, t)-B^D(x, t)|\le M(c^{-3}+H(t)),
\end{equation}
for $x\in\R^3$ and $t\in [0, T]$. It remains to estimate $h=f-f^D$.
Using (\ref{RVMC}), (\ref{DefDarwin}), (\ref{VP}), and (\ref{LVP}),
it is found that
\begin{eqnarray*}
   \lefteqn{\partial_t h+\hat{v}\cdot\nabla_x h+(E+c^{-1}\hat{v}\times B)\cdot\nabla_v h}
   \\ & = & -\partial_t f^D-\hat{v}\cdot\nabla_x f^D-(E+c^{-1}\hat{v}\times B)\cdot\nabla_v f^D
   \\ & = & \Big(v-\frac{1}{2}\,c^{-2}v^2\,v-\hat{v}\Big)\cdot\nabla_x f_0
   +c^{-2}(v-\hat{v})\cdot\nabla_x f_2
   \\ & & +(E^D-E)\cdot\nabla_v f_0
   +c^{-2}(E^D-E)\cdot\nabla_v f_2-c^{-4}E_2\cdot\nabla_v f_2
   \\ & & +c^{-2}((v-\hat{v})\times B_1)\cdot\nabla_v f_0
   +c^{-1}(\hat{v}\times (B^D-B))\cdot\nabla_v f_0-c^{-3}(\hat{v}\times B)\cdot\nabla_v f_2.
\end{eqnarray*}
If $|v|\le M_1$, then also $|\hat{v}|=(1+c^{-2} v^2)^{-1/2}|v|\le |v|\le M_1$ uniformly in $c$,
and hence
\[ \Big|\hat{v}-\Big(1-\frac{1}{2}\,c^{-2}v^2\Big)v\Big|\le Mc^{-4}. \]
Next we note the straightforward estimate $|B_1(x, t)|\le M$ for $|x|\le M_2$
and $t\in [0, T]$, with $B_1$ from (\ref{DefBe}). In view of the bounds (\ref{SchrankeGeschwVP}),
(\ref{SchrankeLVP}), and (\ref{SchrankeFelder}), thus by (\ref{abschE}) and (\ref{abschB}),
\begin{eqnarray}\label{h-vlas}
   & & |\partial_t h(x, v, t)+\hat{v}\cdot\nabla_x h(x, v, t)+(E(x, t)
   +c^{-1}\hat{v}\times B(x, t))\cdot\nabla_v h(x, v, t)|
   \nonumber \\ & & \qquad\qquad\le M(c^{-3}+H(t))
\end{eqnarray}
for $|x|\le M_2$, $|v|\le M_1$, and $t\in [0, T]$. But in $\{(x, v, t): |x|>M_2\}
\cup\{(x, v, t): |v|>M_1\}$ we have $h=f-f^D=0$ by the above definition of $M_1>0$ and $M_2>0$.
Accordingly, (\ref{h-vlas}) is satisfied for all $x\in\R^3$, $v\in\R^3$, and $t\in [0, T]$.
Since $h(x, v, 0)=0$, the argument from \cite[p.~416]{schaeffer:86} yields
\[ H(t)\le\int_0^t M(c^{-3}+H(s))\,ds,  \]
and therefore $H(t)\le Mc^{-3}$ for $t\in [0, T]$. Then due to (\ref{abschE})
and (\ref{abschB}), $|E(x, t)-E^D(x, t)|\le M_R\,c^{-3}$ for $|x|\le R$ and $t\in [0, T]$,
as well as $|B(x, t)-B^D(x, t)|\le Mc^{-3}$ for $x\in\R^3$ and $t\in [0, T]$.
This completes the proof of Theorem \ref{Hauptsatz}. {\hfill$\Box$}\bigskip


\setcounter{equation}{0}

\section{Proof of Theorem \ref{DVM-thm}}
\label{DVM-bew}

In this section we will be sketchy and omit many details,
since the proof is more or less a repetition of what has been said before.
First let us assume that there is a $C^2$-solution $(f^\ast, E^\ast, B^\ast)$
of (\ref{DaVlaMax}), existing on a time interval $[0, T^\ast]$ for some $T^\ast>0$,
such that ${\rm supp}\,f^\ast(\cdot, \cdot, t)\subset\R^3\times\R^3$ is compact
for all $t\in [0, T^\ast]$. Then
\begin{eqnarray}
   B^\ast(x, t) & = & c^{-1}\int |z|^{-2}\bar{z}\times j^\ast(x+z, t)\,dz, \label{Bast-eq} \\
   \Delta E^\ast(x, t) & = & 4\pi\nabla\rho^\ast(x, t)
   +c^{-1}\partial_t\curl B^\ast(x, t). \label{East-eq}
\end{eqnarray}
Since $f^\ast(x, v, 0)=f^\circ(x, v)$ and $j^\ast(x, 0)
=\int (1-\frac{1}{2}\,c^{-2}v^2)v f^\ast(x, v, 0)\,dv
=\int (1-\frac{1}{2}\,c^{-2}v^2)v f^\circ(x, v)\,dv$, it follows that $B^\ast(x, 0)$ 
is determined by $f^\circ$. In order to compute the Poisson integral for $E^\ast$, 
we calculate by means of the transformation $y=w-z$, $dy=dw$, 
and using (\ref{formelv}) below,
\begin{eqnarray*}
   \lefteqn{c^{-1}\Delta^{-1}(\partial_t\curl B^\ast)(x, t)}
   \\ & = & -\frac{1}{4\pi c}\int\frac{dy}{|x-y|}\,\curl\partial_t B^\ast(y, t)
   \\ & = & -\frac{1}{4\pi c^2}\int\frac{dy}{|x-y|}\,\nabla_y\times
   \bigg(\int |z|^{-2}\bar{z}\times\partial_t j^\ast(y+z, t)\,dz\bigg)
   \\ & = & -\frac{1}{4\pi c^2}\int dw\int dz\,|z|^{-2}|x-w+z|^{-1}
   \Big(\bar{z}\,\div (\partial_t j^\ast)(w, t)
   -(\bar{z}\cdot\nabla)\partial_t j^\ast(w, t)\Big)
   \\ & = & \frac{1}{2c^2}\int\frac{dw}{|x-w|}
   \Big([x-w]\,\div (\partial_t j^\ast)(w, t)
   -([x-w]\cdot\nabla)\partial_t j^\ast(w, t)\Big)
   \\ & = & -\frac{1}{2c^2}\int\frac{dz}{|z|}
   \,(1+\bar{z}\otimes\bar{z})\partial_t j^\ast(x+z, t).
\end{eqnarray*}
If we invoke the Vlasov equation for $f^\ast$ and integrate by parts,
this can be rewritten as
\begin{eqnarray*}
   \lefteqn{c^{-1}\Delta^{-1}(\partial_t\curl B^\ast)(x, t)}
   \\ & = & -\frac{1}{2c^2}\int\frac{dz}{|z|}\,(1+\bar{z}\otimes\bar{z})
   \int\Big(1-\frac{1}{2}\,c^{-2}v^2\Big)v\,\partial_t f^\ast(x+z, v, t)\,dv
   \\ & = & \frac{1}{2c^2}\int\frac{dz}{|z|^2}\,\bar{z}\int (3(\bar{z}\cdot v)^2-v^2)
   \Big(1-\frac{1}{2}\,c^{-2}v^2\Big)f^\ast(x+z, v, t)\,dv
   \\ & & -\frac{1}{2c^2}\int\frac{dz}{|z|}\,(1+\bar{z}\otimes\bar{z})
   \int ((E^\ast+c^{-1}v\times B^\ast) f^\ast)(x+z, v, t)\,dv
   \\ & & +\frac{1}{4c^4}\int\frac{dz}{|z|}\,(1+\bar{z}\otimes\bar{z})
   \int v^2 v\,\partial_t f^\ast(x+z, v, t)\,dv. 
\end{eqnarray*}
Therefore the solution $E^\ast$ of (\ref{East-eq}) has the representation
\begin{eqnarray}\label{East-repres}
   E^\ast(x, t) & = & 4\pi\Delta^{-1}(\nabla\rho^\ast)(x, t)
   +c^{-1}\Delta^{-1}(\partial_t\curl B^\ast)(x, t)
   \nonumber \\ & = & -\int |z|^{-2}\bar{z}\,\rho^\ast(x+z, t)\,dz
   \nonumber \\ & & +\frac{1}{2c^2}\int\int |z|^{-2}\bar{z}\,(3(\bar{z}\cdot v)^2-v^2)
   \Big(1-\frac{1}{2}\,c^{-2}v^2\Big)f^\ast(x+z, v, t)\,dv\,dz
   \nonumber \\ & & -\frac{1}{2c^2}\int\int |z|^{-1}\,(1+\bar{z}\otimes\bar{z})
   \int ((E^\ast+c^{-1}v\times B^\ast) f^\ast)(x+z, v, t)\,dv\,dz
   \nonumber \\ & & +\frac{1}{4c^4}\int\int |z|^{-1}v^2
   (v+(\bar{z}\cdot v)\bar{z})\,\partial_t f^\ast(x+z, v, t)\,dv\,dz. 
\end{eqnarray}
Comparison with (\ref{VP}) and (\ref{ADarEz}) reveals the analogy
to $E^D$ at the relevant orders of $c^{-1}$. In particular, 
if we evaluate this relation at $t=0$, the Banach fixed point theorem 
applied in $C_b(\R^3)$ shows that for $c\ge c^\ast$ sufficiently large 
the function $E^\ast(x, 0)$ is uniquely determined 
by $f^\circ(x, v)=f^\ast(x, v, 0)$. Thus $f^\circ$ alone already fixes 
$E^\circ$ and $B^\circ$. Concerning the local and uniform (in $c$) existence 
of a solution to (\ref{DaVlaMax}) and the conservation of energy, 
one can use (\ref{Bast-eq}) and (\ref{East-repres}) to follow 
the usual method by setting up an iteration scheme for which convergence 
can be verified on a small time interval; cf.~\cite[Sect.~5.8]{glassey:96}. 
Finally, by similar arguments as used in the proof of Theorem \ref{Hauptsatz}
it can be shown that solutions of (\ref{DaVlaMax}) approximate
solutions of (\ref{RVMC}) up to an error of order $c^{-3}$.
{\hfill$\Box$}\bigskip


\setcounter{equation}{0}

\section{Appendix}
\label{append}

\subsection{Representation Formulas}

\subsubsection{Representation of the approximation fields $E^D$ and $B^D$}
\label{repapp-sect}

Here we will derive the representation formula (\ref{DarstellungED})
for the approximate field $E^D$ from (\ref{DefDarwin}).
Since the calculations for the electric and the magnetic field
are quite similar, we will only analyze in detail the electric field
and simply state the result for its magnetic counterpart.

From (\ref{DefDarwin}) we recall $E^D=E_0+c^{-2}E_2$, where
\begin{eqnarray}
   E_0(x, t) & = & -\int |z|^{-2}\bar{z}\,\rho_0(x+z, t)\,dz, \label{ma1} \\
   E_2(x, t) & = & \frac{1}{2}\int\bar{z}\,\partial_t^2\rho_0(x+z, t)\,dz
   -\int |z|^{-1}\partial_t j_0(x+z, t)\,dz
   -\int |z|^{-2}\bar{z}\,\rho_2(x+z, t)\,dz,\qquad\label{ma2}
\end{eqnarray}
cf.~(\ref{VP}) and (\ref{DefEz}). We split the domain of integration
in $\{|z|>ct\}$ and $\{|z|\leq ct\}$, and to handle the interior part
$\{|z|\leq ct\}$ we expand the densities w.r.t.~$t$ about the retarded time
\[ \hat{t}(z):=t-c^{-1}|z|. \]
To begin with, we have
\begin{eqnarray}\label{EDI1}
   \lefteqn{-\int_{|z|\leq ct}|z|^{-2}\bar{z}\,\rho_0(x+z, t)\,dz
   =-\int_{|z|\leq ct}|z|^{-2}\bar{z}\,\rho_0(x+z, \hat{t}(z))\,dz} \nonumber \\
   & & -c^{-1}\int_{|z|\leq ct}|z|^{-1}\bar{z}\,\partial_t\rho_0(x+z, \hat{t}(z))\,dz
   -c^{-2}\frac{1}{2}\int_{|z|\leq ct}\bar{z}\,\partial_t^2\rho_0(x+z, \hat{t}(z))\,dz
   \nonumber\\
   & & -\frac{1}{2}\int_{|z|\leq ct}|z|^{-2}\bar{z}\int_{\widehat t (z)}^t
   (t-s)^2\partial_t^3\rho_0(x+z, s)\,ds\,dz.
\end{eqnarray}
Using (\ref{SchrankeGeschwVP}) and (\ref{SchrankeFeldVP}), the last term
is ${\cal O}_{cpt}(c^{-3})$; note that $|x|\le R$ for some $R>0$ together
with the support properties of $f_0$ imply that we only have
to integrate in $z$ over a set which is uniformly bounded in $c\ge 1$.
Since $\partial_t\rho_0+\div j_0=0$ by (\ref{VP}), we also find
\begin{eqnarray}\label{EDI2}
   \lefteqn{-c^{-1}\int_{|z|\leq ct}|z|^{-1}\bar{z}\,\partial_t\rho_0(x+z, \hat{t}(z))\,dz
   =c^{-1}\int_{|z|\leq ct}|z|^{-1}\bar{z}\,\nabla_x\cdot j_0(x+z, \hat{t}(z))\,dz}
   \nonumber \\ & = & c^{-1}\int_{|z|\leq ct}\int |z|^{-1}\bar{z}
   \,v\cdot\nabla_x f_0(x+z, v, \hat{t}(z))\,dv\,dz \nonumber\\
   & = & c^{-1}\int_{|z|\leq ct}\int |z|^{-1}\bar{z}\,v\cdot
   \Big(\nabla_z [f_0(x+z, v, \hat{t}(z))]
   +c^{-1}\bar{z}\,\partial_t f_0(x+z, v, \hat{t}(z))\Big)\,dv\,dz\nonumber\\
   & = & I+II,
\end{eqnarray}
with
\begin{eqnarray}\label{EDI3}
   I & = & c^{-1}\int_{|z|\leq ct}\int |z|^{-1}\bar{z}\,
   v\cdot\nabla_z [f_0(x+z, v, \hat{t}(z))]\,dv\,dz
   \nonumber\\
   & = & -c^{-1}\int_{|z|\leq ct}\int\left(\nabla_z\cdot
   \left[|z|^{-1}\bar{z}_iv\right]\right)_{i=1,2,3}f_0(x+z, v, \hat{t}(z))\,dv\,dz
   \nonumber\\
   & & +c^{-1}(ct)^{-1}\int_{|z|=ct}\int\bar{z}(\bar{z}\cdot v) f^\circ(x+z, v)\,dv\,ds(z)
   \nonumber\\
   & = & -c^{-1}\int_{|z|\leq ct}\int |z|^{-2}(v-2(\bar{z}\cdot v)\bar{z})
   f_0(x+z, v, \hat{t}(z))\,dv\,dz\nonumber\\
   & & +c^{-1}(ct)^{-1}\int_{|z|=ct}\int\bar{z}(\bar{z}\cdot v)f^\circ(x+z, v)\,dv\,ds(z);
\end{eqnarray}
observe that $\hat{t}(z)=0$ for $|z|=ct$ was used for the boundary term.
Similarly, by (\ref{VP}),
\begin{eqnarray}\label{EDI4}
   II & = & c^{-2}\int_{|z|\leq ct}\int |z|^{-1}\bar{z} (\bar{z}\cdot v)
   \,\partial_t f_0(x+z, v, \hat{t}(z))\,dv\,dz
   \nonumber\\
   & = & -c^{-2}\int_{|z|\leq ct}\int |z|^{-1}\bar{z} (\bar{z}\cdot v)
   (v\cdot\nabla_x f_0+E_0\cdot\nabla_v f_0)(x+z, v, \hat{t}(z))\,dv\,dz
   \nonumber\\
   & = & c^{-2}\int_{|z|\leq ct}\int\nabla_z\cdot {(|z|^{-1}\bar{z}_i
   (\bar{z}\cdot v)v)}_{i=1,2,3}\,f_0(x+z, v, \hat{t}(z))\,dv\,dz
   \nonumber\\
   & & -c^{-2}(ct)^{-1}\int_{|z|=ct}\int\bar{z}(\bar{z}\cdot v)^2 f^\circ(x+z, v)
   \,dv\,ds(z)
   \nonumber\\
   & & -c^{-3}\int_{|z|\leq ct}\int |z|^{-1}\bar{z}(\bar{z}\cdot v)^2
   \partial_t f_0(x+z, v, \hat{t}(z))\,dv\,dz
   \nonumber\\
   & & +c^{-2}\int_{|z|\leq ct}\int |z|^{-1}\bar{z}\,\bar{z}\cdot E_0 f_0(x+z, v, \hat{t}(z))
   \,dv\,dz
   \nonumber\\
   & = & c^{-2}\int_{|z|\leq ct}\int |z|^{-2}((\bar{z}\cdot v)v
   +v^2\bar{z}-3\bar{z}(\bar{z}\cdot v)^2)\,f_0(x+z, v, \hat{t}(z))\,dv\,dz \nonumber\\
   & & +c^{-2}\int_{|z|\leq ct}|z|^{-1}\bar{z}\,\bar{z}\cdot E_0\rho_0(x+z, \hat{t}(z))\,dz
   \nonumber\\
   & & -c^{-2}(ct)^{-1}\int_{|z|=ct}\int\bar{z}(\bar{z}\cdot v)^2 f^\circ(x+z, v)\,dv\,ds(z)
   +{\cal O}_{cpt}(c^{-3}).
\end{eqnarray}
Next, due to (\ref{SchrankeGeschwVP}) and (\ref{SchrankeFeldVP}) we also have
\begin{equation}\label{EDI4a}
   -c^{-2}\frac{1}{2}\int_{|z|\leq ct}\bar{z}\,\partial_t^2\rho_0(x+z, \hat{t}(z))\,dz
   =-c^{-2}\frac{1}{2}\int_{|z|\leq ct}\bar{z}\,\partial_t^2\rho_0(x+z, t)\,dz
   +{\cal O}_{cpt}(c^{-3}).
\end{equation}
Thus so far by (\ref{ma1}) and (\ref{EDI1})--(\ref{EDI4a}),
\begin{eqnarray}\label{ma3}
   E_0(x, t) & = & -\int_{|z|>ct}|z|^{-2}\bar{z}\,\rho_0(x+z, t)\,dz
   -\int_{|z|\leq ct}|z|^{-2}\bar{z}\,\rho_0(x+z, t)\,dz
   \nonumber\\ & = & -\int_{|z|>ct}|z|^{-2}\bar{z}\,\rho_0(x+z, t)\,dz
   \nonumber\\ & & -\int_{|z|\leq ct}|z|^{-2}\bar{z}\,\rho_0(x+z, \hat{t}(z))\,dz
   -c^{-2}\frac{1}{2}\int_{|z|\leq ct}\bar{z}\,\partial_t^2\rho_0(x+z, t)\,dz
   \nonumber\\
   & & -c^{-1}\int_{|z|\leq ct}\int |z|^{-2}(v-2(\bar{z}\cdot v)\bar{z})
   f_0(x+z, v, \hat{t}(z))\,dv\,dz
   \nonumber\\
   & & +c^{-2}\int_{|z|\leq ct}\int |z|^{-2}((\bar{z}\cdot v)v
   +v^2\bar{z}-3\bar{z}(\bar{z}\cdot v)^2)\,f_0(x+z, v, \hat{t}(z))\,dv\,dz \nonumber\\
   & & +c^{-2}\int_{|z|\leq ct}|z|^{-1}\bar{z}\,\bar{z}\cdot E_0\rho_0(x+z, \hat{t}(z))\,dz
   \nonumber\\
   & & +c^{-1}(ct)^{-1}\int_{|z|=ct}\int (\bar{z}\cdot v)\bar{z}\,f^\circ(x+z, v)\,dv\,ds(z)
   \nonumber \\
   & & -c^{-2}(ct)^{-1}\int_{|z|=ct}\int (\bar{z}\cdot v)^2\bar{z}\,f^\circ(x+z, v)\,dv\,ds(z)
   +{\cal O}_{cpt}(c^{-3}).
\end{eqnarray}
Now we turn to $E_2$, cf.~(\ref{ma2}). Since $E_2$ enters $E^D$ with the factor $c^{-2}$,
we first note that
\[ c^{-2}\frac{1}{2}\int_{|z|\leq ct}\bar{z}\,\partial_t^2\rho_0(x+z, t)\,dz \]
cancels a term on the right-hand side of (\ref{ma3}). In addition, by analogous arguments,
\begin{eqnarray}\label{EDI5}
   \lefteqn{-c^{-2}\int_{|z|\leq ct} |z|^{-1}\partial_t j_0(x+z, t)\,dz}
   \nonumber \\
   & = & -c^{-2}\int_{|z|\leq ct} |z|^{-1}\partial_t j_0(x+z, \hat{t}(z))\,dz
   -c^{-2}\int_{|z|\leq ct} |z|^{-1}\int_{\hat{t}(z)}^t\partial_t^2 j_0(x+z, s)\,ds\,dz
   \nonumber \\
   & = & -c^{-2}\int_{|z|\leq ct} |z|^{-1}\int v\,\partial_t f_0(x+z, v, \hat{t}(z))\,dv\,dz
   +{\cal O}_{cpt}(c^{-3}) \nonumber\\
   & = & c^{-2}\int_{|z|\leq ct} |z|^{-1}\int v\,(v\cdot\nabla_x f_0+E_0\cdot\nabla_v f_0)
   (x+z, v, \hat{t}(z))\,dv\,dz+{\cal O}_{cpt}(c^{-3})
   \nonumber \\
   & = & c^{-2}\int_{|z|\leq ct} |z|^{-1}\int v\,v\cdot\Big(\nabla_z [f_0(x+z, v, \hat{t}(z))]
   +c^{-1}\bar{z}\,\partial_t f_0(x+z, v, \hat{t}(z))\Big)\,dv\,dz\nonumber\\
   & & +c^{-2}\int_{|z|\leq ct} |z|^{-1}\int v\,E_0\cdot\nabla_v f_0(x+z, v, \hat{t}(z))\,dv\,dz
   +{\cal O}_{cpt}(c^{-3})\nonumber \\
   & = & c^{-2}\int_{|z|\leq ct}\int |z|^{-2}(\bar{z}\cdot v)v f_0(x+z, v, \hat{t}(z))\,dv\,dz
   \nonumber\\
   & & +c^{-2}(ct)^{-1}\int_{|z|=ct}\int (\bar{z}\cdot v)v f^\circ(x+z, v)\,dv\,ds(z)
   \nonumber\\
   & & -c^{-2}\int_{|z|\leq ct} |z|^{-1} (E_0\rho_0)(x+z, \hat{t}(z))\,dz+{\cal O}_{cpt}(c^{-3}).
\end{eqnarray}
Finally,
\begin{equation}\label{ma4}
   -c^{-2}\int_{|z|\le ct} |z|^{-2}\bar{z}\,\rho_2(x+z, t)\,dz
   =-c^{-2}\int_{|z|\le ct} |z|^{-2}\bar{z}\,\rho_2(x+z, \hat{t}(z))\,dz
   +{\cal O}_{cpt}(c^{-3}).
\end{equation}
Therefore if we write
\begin{eqnarray*}
   c^{-2}E_2(x, t) & = & c^{-2}\frac{1}{2}\int_{|z|>ct}\bar{z}\,\partial_t^2\rho_0(x+z, t)\,dz
   -c^{-2}\int_{|z|>ct} |z|^{-1}\partial_t j_0(x+z, t)\,dz
   \\ & & -c^{-2}\int_{|z|>ct} |z|^{-2}\bar{z}\,\rho_2(x+z, t)\,dz
   +c^{-2}\frac{1}{2}\int_{|z|\leq ct}\bar{z}\,\partial_t^2\rho_0(x+z, t)\,dz
   \\ & & -c^{-2}\int_{|z|\leq ct} |z|^{-1}\partial_t j_0(x+z, t)\,dz
   -c^{-2}\int_{|z|\leq ct} |z|^{-2}\bar{z}\,\rho_2(x+z, t)\,dz,
\end{eqnarray*}
use (\ref{EDI5}) and (\ref{ma4}), and thereafter add the result to (\ref{ma3}),
it turns out that $E^D=E_0+c^{-2}E_2$ can be decomposed as claimed in (\ref{DarstellungED}).

Similar calculations for $B^D(x, t)=c^{-1}B_1(x, t)$ using (\ref{DefBe}) yield
\begin{eqnarray}\label{RepBD}
   B^D(x, t) & = & c^{-1}\int_{|z|>ct} |z|^{-2}\bar{z}\times j_0(x+z, t)\,dz
   \nonumber \\
   & & +c^{-1}\int_{|z|\leq ct} |z|^{-2}\bar{z}\times j_0(x+z, \hat{t}(z))\,dz
   \nonumber \\
   & & -2c^{-2}\int_{|z|\leq ct}\int |z|^{-2}(\bar{z}\cdot v)(\bar{z}\times v)
   \,f_0(x+z, v, \hat{t}(z))\,dv\,dz \nonumber\\
   & & +c^{-2}\int_{|z|\leq ct} |z|^{-1}\bar{z}\times E_0\rho_0(x+z, \hat{t}(z))\,dz
   \nonumber\\
   & & -c^{-2}(ct)^{-1}\int_{|z|=ct}\int (\bar{z}\cdot v)(\bar{z}\times v)
   \,f^\circ(x+z, v)\,dv\,ds(z)+{\cal O}(c^{-3}).
\end{eqnarray}

\subsubsection{Representation of the Maxwell fields $E$ and $B$}
\label{max-repres}

In this section we will verify the representation formula (\ref{DarstellungE})
for the full Maxwell field $E$, by expanding the respective expressions
from \cite{glstr,schaeffer:86} to higher orders. Once again
the computation for the corresponding magnetic field $B$ is very similar
and therefore omitted. Let $(f, E, B)$ be a $C^1$-solution
of (\ref{RVMC}) with initial data $(f^\circ, E^\circ, B^\circ)$. We recall the following
representation from \cite[(A13), (A14), (A3)]{schaeffer:86}:
\begin{eqnarray}
   E & = & E_D+E_{DT}+E_T+E_S, \label{EFeld} \\
   B & = & B_D+B_{DT}+B_T+B_S, \label{BFeld}
\end{eqnarray}
where
\begin{eqnarray*}
   E_D(x, t) & = &
   \partial_t\bigg(\frac{t}{4\pi}\int_{|\omega|=1} E^\circ(x+ct\omega)\,d\omega\bigg)
   +\frac{t}{4\pi}\int_{|\omega|=1}\partial_t E(x+ct\omega, 0)\,d\omega, \\
   E_{DT}(x, t) & = & -(ct)^{-1}\int_{|z|=ct}\int K_{DT}(\bar{z}, \hat{v})
   f^\circ(x+z, v)\,dv\,ds(z), \\
   E_T(x, t) & = & -\int_{|z|\leq ct} |z|^{-2}\int K_T(\bar{z}, \hat{v})
   f(x+z, v, \hat{t}(z))\,dv\,dz, \\
   E_S(x, t) & = &-c^{-2}\int_{|z|\leq ct} |z|^{-1}\int K_S(\bar{z}, \hat{v})
   (E+c^{-1}\hat{v}\times B)f(x+z, v, \hat{t}(z))\,dv\,dz,
\end{eqnarray*}
and
\begin{eqnarray*}
   B_D(x, t) & = &
   \partial_t\bigg(\frac{t}{4\pi}\int_{|\omega|=1} B^\circ(x+ct\omega)\,d\omega\bigg)
   +\frac{t}{4\pi}\int_{|\omega|=1}\partial_t B(x+ct\omega, 0)\,d\omega, \\
   B_{DT}(x, t) & = & (ct)^{-1}\int_{|z|=ct}\int L_{DT}(\bar{z}, \hat{v})
   f^\circ(x+z, v)\,dv\,ds(z), \\
   B_T(x, t) & = & c^{-1}\int_{|z|\leq ct}
   |z|^{-2}\int L_T(\bar{z}, \hat{v}) f(x+z, v, \hat{t}(z))\,dv\,dz, \\
   B_S(x, t) & = & c^{-2}\int_{|z|\leq ct} |z|^{-1}\int L_S(\bar{z}, \hat{v})
   (E+c^{-1}\hat{v}\times B)f(x+z, v, \hat{t}(z))\,dv\,dz,
\end{eqnarray*}
with $\bar{z}=|z|^{-1}z$ and $\hat{t}(z)=t-c^{-1}|z|$. The kernels are given by
\begin{eqnarray*}
   K_{DT}(\bar{z}, \hat{v}) & = & (1+c^{-1}\bar{z}\cdot\hat{v})^{-1}
   (\bar{z}-c^{-2}(\bar{z}\cdot\hat{v})\hat{v}), \\
   K_T(\bar{z}, \hat{v})  & = & (1+c^{-1}\bar{z}\cdot\hat{v})^{-2}(1-c^{-2}\hat{v}^2)
   (\bar{z}+c^{-1}\hat{v}), \\
   K_S(\bar{z}, \hat{v})  & = & (1+c^{-1}\bar{z}\cdot\hat{v})^{-2}(1+c^{-2} v^2)^{-1/2}
   \Big[(1+c^{-1}\bar{z}\cdot\hat{v})+c^{-2}((\bar{z}\cdot\hat{v})\bar{z}-\hat{v})\otimes\hat{v}
   \\ & & \hspace{14em} -(\bar{z}+c^{-1}\hat{v})\otimes\bar{z}\Big]\in\R^{3\times 3},
\end{eqnarray*}
and
\begin{eqnarray*}
   L_{DT}(\bar{z}, \hat{v}) & = & (1+c^{-1}\bar{z}\cdot\hat{v})^{-1}
   (\bar{z}\times c^{-1}\hat{v}), \\
   L_T(\bar{z}, \hat{v})  & = & (1+c^{-1}\bar{z}\cdot\hat{v})^{-2}
   (1-c^{-2}\hat{v}^2)(\bar{z}\times\hat{v}), \\
   L_S(\bar{z}, \hat{v})  & = & (1+c^{-1}\bar{z}\cdot\hat{v})^{-2}(1+c^{-2}v^2)^{-1/2}
   \Big[(1+c^{-1}\bar{z}\cdot\hat{v})\bar{z}\times (\ldots)
   \\ & & \hspace{14em} -c^{-2}(\bar{z}\times\hat{v})\otimes 
   (c\bar{z}+\hat{v})\Big]\in\R^{3\times 3}.
\end{eqnarray*}
Next we expand these fields in powers of $c^{-1}$.
According to (\ref{SchrankeSupport}) we can assume that the
$v$-support of $f(x, \cdot, t)$ is uniformly bounded in $x\in\R^3$
and $t\in [0, T]$, say $f(x, v, t)=0$ for $|v|\ge P:=\max_{t\in
[0, T]}P_{V\!M}(t)$. Thus we may suppose that $|v|\le P$ in each
of the $v$-integrals, and hence also $|\hat{v}|=(1+c^{-2}
v^2)^{-1/2}|v|\le |v|\le P$ uniformly in $c$. It follows that
\[ \hat{v}=\Big(1-\frac{1}{2}\,c^{-2}v^2\Big)v+{\cal O}(c^{-4}). \]
For instance, for the kernel $K_{DT}$ of $E_{DT}$ this yields
\begin{eqnarray*}
   K_{DT}(\bar{z}, \hat{v}) & = & (1+c^{-1}\bar{z}\cdot\hat{v})^{-1}
   (\bar{z}-c^{-2}(\bar{z}\cdot\hat{v})\hat{v})
   \\ & = & \Big(1-c^{-1}\bar{z}\cdot v+c^{-2}(\bar{z}\cdot v)^2+{\cal O}(c^{-3})\Big)
   \Big(\bar{z}-c^{-2}(\bar{z}\cdot v)v+{\cal O}(c^{-4})\Big)
   \\ & = & \bar{z}-c^{-1}(\bar{z}\cdot v)\bar{z}+c^{-2}(\bar{z}\cdot v)^2\bar{z}
   -c^{-2}(\bar{z}\cdot v)v+{\cal O}(c^{-3}).
\end{eqnarray*}
If we choose $R_0>0$ such that $f^\circ(x, v)=0$ for $|x|\ge R_0$, then
\begin{eqnarray*}
   -(ct)^{-1}\int_{|z|=ct}\int_{|v|\le P} {\cal O}(c^{-3})
   {\bf 1}_{B(0, R_0)}(x+z)\,dv\,ds(z)
   & = & \bigg(ct\int_{|\omega|=1}{\bf 1}_{B(0, R_0)}(x+ct\omega)\,ds(\omega)\bigg){\cal O}(c^{-3})
   \\ & = & {\cal O}(c^{-3})
\end{eqnarray*}
by \cite[Lemma 1]{schaeffer:86}, uniformly in $x\in\R^3$, $t\in [0, T]$, and $c\ge 1$.
Therefore we arrive at
\begin{eqnarray}\label{EDT-expa}
   E_{DT}(x, t) & = & -(ct)^{-1}\int_{|z|=ct}\int
  \Big(\bar{z}-c^{-1}(\bar{z}\cdot v)\bar{z}+c^{-2}[(\bar{z}\cdot v)^2\bar{z}
   -(\bar{z}\cdot v)v]\Big) f^\circ(x+z, v)\,dv\,ds(z)
  \nonumber \\ & & +{\cal O}(c^{-3}).
\end{eqnarray}
Concerning $E_T$, we note that $f(x, v, t)=0$ for $|x|\ge R_0+TP=:R_1$.
Since, by distinguishing the cases $|x-y|\ge 1$ and $|x-y|\le 1$,
\[ \int_{|z|\leq ct} |z|^{-2}\,{\bf 1}_{B(0, R_1)}(x+z)\,dz
   =\int_{|x-y|\leq ct} |x-y|^{-2}\,{\bf 1}_{B(0, R_1)}(y)\,dy={\cal O}(1) \]
uniformly in $x\in\R^3$, $t\in [0, T]$, and $c\ge 1$, similar computations
as before show that
\begin{eqnarray}\label{ET-expa}
   E_T(x, t) & = & -\int_{|z|\leq ct} |z|^{-2}\int\Big(\bar{z}
   +c^{-1}[v-2(\bar{z}\cdot v)\bar{z}]+c^{-2}[3(\bar{z}\cdot v)^2\bar{z}-v^2\bar{z}
   -2(\bar{z}\cdot v)v]\Big) \nonumber \\
   & & \hspace{8em} f(x+z, v, \hat{t}(z))\,dv\,dz+{\cal O}(c^{-3}).
\end{eqnarray}
In the same manner, elementary calculations using also (\ref{SchrankeFelder})
can be carried out to get
\begin{eqnarray}
   E_S(x, t) & = & -c^{-2}\int_{|z|\leq ct} |z|^{-1}\int (1-\bar{z}\otimes\bar{z})
   (Ef)(x+z, v, \hat{t}(z))\,dv\,dz+{\cal O}(c^{-3}), \label{ES-expa} \\
   B_{DT}(x, t) & = & (ct)^{-1}\int_{|z|=ct}\int\Big(c^{-1}\bar{z}\times v
   -c^{-2}(\bar{z}\cdot v)\bar{z}\times v\Big)f^\circ(x+z, v)\,dv\,ds(z)
   \nonumber \\ & & +{\cal O}(c^{-3}), \label{BDT-expa}
   \\ B_T(x, t) & = & c^{-1}\int_{|z|\leq ct}
   |z|^{-2}\int (\bar{z}\times v-c^{-1}2v\cdot \bar{z}\bar{z}\times v)
   f(x+z, v, \hat{t}(z))\,dv\,dz+{\cal O}(c^{-3}),\qquad\label{BT-expa} \\
   B_S(x, t) & = & c^{-2}\int_{|z|\leq ct} |z|^{-1}\int\bar{z}\times (Ef)(x+z, v, \hat{t}(z))
   \,dv\,dz+{\cal O}(c^{-3}). \label{BS-expa}
\end{eqnarray}
Next we consider the data term
\begin{eqnarray}\label{ma5}
   E_D(x, t) & = &
   \partial_t\bigg(\frac{t}{4\pi}\int_{|\omega|=1} E^\circ(x+ct\omega)\,d\omega\bigg)
   +\frac{t}{4\pi}\int_{|\omega|=1}\partial_t E(x+ct\omega, 0)\,d\omega,
   \nonumber \\ & =: & III+IV.
\end{eqnarray}
Since $f_2(x, v, 0)=0$ by (\ref{LVP}), we have $\rho_2(x, 0)=0$.
Thus we get from (\ref{IC}), (\ref{VP}), and (\ref{DefEz}),
\begin{eqnarray*}
   E^\circ(x) & = & E_0(x, 0)+c^{-2}E_2(x, 0)
   \\ & = & -\int |z|^{-2}\bar{z}\,\rho_0(x+z, 0)\,dz
   +c^{-2}\bigg(\frac{1}{2}\int\bar{z}\,\partial_t^2\rho_0(x+z, 0)\,dz
   -\int |z|^{-1}\partial_t j_0(x+z, 0)\,dz\bigg).
\end{eqnarray*}
Using the formulas (\ref{formelii}), (\ref{formeliv}), and (\ref{formeli}) below, we calculate
\begin{eqnarray*}
   \lefteqn{-\int_{|\omega|=1}\int |z|^{-2}\bar{z}\,\rho_0(x+ct\omega+z, 0)\,dz\,d\omega}
   \hspace{2em} \\
   & = & -\int\rho_0(y,0)\int_{|\omega|=1}|y-x-ct\omega|^{-3}(y-x-ct\omega)\,d\omega\,dy \\
   & = & -4\pi\int_{|z|>ct} |z|^{-2}\bar{z}\,\rho_0(x+z, 0)\,dz, \\
   \lefteqn{\int_{|\omega|=1}\int\bar{z}\,\partial_t^2\rho_0(x+ct\omega+z, 0)\,dz\,d\omega}
   \hspace{2em} \\
   & = & \int\partial_t^2\rho_0(y,0)\int_{|\omega|=1}|y-x-ct\omega|^{-1}(y-x-ct\omega)
   \,d\omega\,dy \\
   & = & 4\pi\int_{|z|>ct}\Big(\bar{z}-\frac{1}{3}(ct)^2|z|^{-2}\bar{z}\Big)
   \,\partial_t^2\rho_0(x+z, 0)\,dz
   +\frac{8\pi}{3ct}\int_{|z|\leq ct}z\,\partial_t^2\rho_0(x+z, 0)\,dz, \\
   \lefteqn{-\int_{|\omega|=1}\int |z|^{-1}\partial_t j_0(x+ct\omega+z, 0)\,dz\,d\omega}
   \hspace{2em} \\
   & = & -\int\partial_t j_0(y,0)\int_{|\omega|=1}|y-x-ct\omega|^{-1}\,d\omega\,dy \\
   & = & -4\pi\int_{|z|>ct}|z|^{-1}\partial_t j_0(x+z, 0)\,dz
   -\frac{4\pi}{ct}\int_{|z|\leq ct}\partial_t j_0(x+z, 0)\,dz.
\end{eqnarray*}
Therefore we get
\begin{eqnarray}\label{III-form}
   III & = & \partial_t\bigg(\frac{t}{4\pi}\int_{|\omega|=1}
   E^\circ(x+ct\omega)\,d\omega\bigg)
   \nonumber \\ & = & \partial_t\bigg(-t\int_{|z|>ct} |z|^{-2}\bar{z}
   \,\rho_0(x+z, 0)\,dz
   +\frac{t}{2c^2}\int_{|z|>ct}\bar{z}\,\partial_t^2\rho_0(x+z, 0)\,dz
   \nonumber \\ & & \hspace{2em} -\frac{t^3}{6}\int_{|z|>ct}|z|^{-2}\bar{z}
   \,\partial_t^2\rho_0(x+z, 0)\,dz
   +\frac{1}{3c^3}\int_{|z|\leq ct}z\,\partial_t^2\rho_0(x+z, 0)\,dz
   \nonumber \\ & & \hspace{2em} -\frac{t}{c^2}\int_{|z|>ct}|z|^{-1}
   \partial_t j_0(x+z, 0)\,dz
   -\frac{1}{c^3}\int_{|z|\leq ct}\partial_t j_0(x+z, 0)\,dz\bigg)
   \nonumber \\ & = & -\int_{|z|>ct} |z|^{-2}\bar{z}\,\rho_0(x+z, 0)\,dz
   +\frac{1}{2}\,c^{-2}\int_{|z|>ct}\bar{z}\,\partial_t^2\rho_0(x+z, 0)\,dz
   \nonumber \\ & & -\frac{1}{2}\,t^2\int_{|z|>ct}|z|^{-2}\bar{z}
   \,\partial_t^2\rho_0(x+z, 0)\,dz
   -c^{-2}\int_{|z|>ct}|z|^{-1}\partial_t j_0(x+z, 0)\,dz
   \nonumber \\ & & +(ct)^{-1}\int_{|z|=ct} \bar{z}\,\rho_0(x+z, 0)\,ds(z),
\end{eqnarray}
note that several terms have cancelled here.
Now we discuss the second part $IV$ of the data term $E_D$, cf.~(\ref{ma5}).
To begin with, by (\ref{IC}), (\ref{DefBe}), and (\ref{VP}),
\begin{eqnarray*}
   B(x, 0) & = & B^\circ(x)=c^{-1}B_1(x, 0)=c^{-1}\int |z|^{-2}\bar{z}\times j_0(x+z, 0)\,dz
   \\ & = & c^{-1}\int\int |z|^{-2}(\bar{z}\times v)f_0(x+z, v, 0)\,dv\,dz,
\end{eqnarray*}
due to $j_0=\int v f_0\,dv$. Therefore using (\ref{VP}) for $f_0$ and integration by parts,
we obtain
\begin{eqnarray*}
   \curl B(x, 0) & = & c^{-1}\curl\int\int |z|^{-2}(\bar{z}\times v)f_0(x+z, v, 0)\,dv\,dz \\
   & = & c^{-1}\int\int |z|^{-2}\,\nabla_x f_0\times(\bar{z}\times v)\,dv\,dz\\
   & = & c^{-1}\int\int |z|^{-2}\Big((v\cdot\nabla_x f_0)\bar{z}
   -(\bar{z}\cdot\nabla_x f_0)v\Big)\,dv\,dz \\
   & = & c^{-1}\int\int |z|^{-2}\bar{z}
   \,(-\partial_t f_0-E_0\cdot\nabla_v f)\,dv\,dz \\
   & & -c^{-1}\int\int |z|^{-2}v\,\bar{z}\cdot\nabla_z [f_0(\ldots)]\,dv\,dz \\
   & = & -c^{-1}\int\int |z|^{-2}\bar{z}\,\partial_t f_0(x+z, v, 0)\,dv\,dz
   +c^{-1}4\pi\int vf_0(x, v, 0)\,dv,
\end{eqnarray*}
by observing that in general for a continuous function $g$,
\[ -\int_{|z|>\eps}|z|^{-2}\,\bar{z}\cdot\nabla_z g(x+z)\,dz
   =\int_{|\omega|=1}g(x+\eps\omega)\,d\omega\to 4\pi g(x),\quad\eps\to 0. \]
From Maxwell's equations we have $\partial_t E(x, t)=c\curl B(x, t)-4\pi j(x, t)$, 
and thus in view of $j(x, 0)=\int\hat{v} f(x, v, 0)\,dv=\int\hat{v} f^\circ(x, v)\,dv
=\int\hat{v} f_0(x, v, 0)\,dv$,
\begin{eqnarray*}
   \partial_t E(x,0) & = & c\curl B(x, 0)-4\pi j(x, 0)
   \\ & = & -\int\int |z|^{-2}\bar{z}\,\partial_t f_0(x+z, v, 0)\,dv\,dz
   +4\pi\int (v-\hat{v})f_0(x, v, 0)\,dv.
\end{eqnarray*}
Hence due to (\ref{formelii}), $v-\hat{v}={\cal O}(c^{-2})$,
and by \cite[Lemma 1]{schaeffer:86},
\begin{eqnarray}\label{IV-form}
   IV & = & \frac{t}{4\pi}\int_{|\omega|=1}\partial_t E(x+ct\omega, 0)\,d\omega
   \nonumber \\ & = & -\frac{t}{4\pi}\int_{|\omega|=1}
   \int\int |z|^{-2}\bar{z}\,\partial_t f_0(x+ct\omega+z, v, 0)\,dv\,dz\,d\omega
   \nonumber \\ & & +\,c^{-1}(ct)\int_{|\omega|=1}\int (v-\hat{v})
   f^\circ(x+ct\omega, v)\,dv\,d\omega
   \nonumber \\ & = & -t\int_{|z|>ct} |z|^{-2}\bar{z}\,\partial_t\rho_0(x+z, 0)\,dz
   +{\cal O}(c^{-3}).
\end{eqnarray}
If we combine (\ref{EFeld}), (\ref{ma5}), (\ref{III-form}),
(\ref{IV-form}), (\ref{EDT-expa}), (\ref{ET-expa}), and (\ref{ES-expa}),
then we see that (\ref{DarstellungE}) is satisfied.

A similar calculation yields
\begin{eqnarray}\label{BD-form}
   B_D(x, t) & = & c^{-1}\int_{|z|>ct} |z|^{-2}\bar{z}\times j_0(x+z,0)\,dz
   +c^{-1} t\int_{|z|>ct} |z|^{-2}\bar{z}\times\partial_t j_0(x+z,0)\,dz \nonumber \\
   & & -c^{-1}(ct)^{-1}\int_{|z|=ct}\bar{z}\times j_0(x+z,0)\,ds(z),
\end{eqnarray}
and an analogous decomposition of $B$ into $B=B_{{\rm ext}}+B_{{\rm int}}
+B_{{\rm bd}} +{\cal O}(c^{-3})$.

\subsection{Some explicit integrals}

We point out some formulas that have been used in the previous sections.
For $z\in\R^3$ and $r>0$ an elementary calculation yields
\begin{equation}\label{formeli}
  \int_{|\omega|=1}|z-r\omega|^{-1}\,d\omega
  =\left\{\begin{array}{c@{\quad:\quad}c}
  4\pi r^{-1} & r\geq |z| \\ 4\pi |z|^{-1} & r\leq |z|
  \end{array}\right. .
\end{equation}
Differentiation w.r.t.~$z$ gives
\begin{equation}\label{formelii}
   \int_{|\omega|=1}|z-r\omega|^{-3}(z-r\omega)\,d\omega
   =\left\{\begin{array}{c@{\quad:\quad}c}
   0 & r>|z| \\ 4\pi |z|^{-2}\bar{z} & r<|z|
   \end{array}\right. .
\end{equation}
Similarly,
\[ \int_{|\omega|=1}|z-r\omega|\,d\omega
   =\left\{\begin{array}{c@{\quad:\quad}c}
   4\pi r+\frac{4\pi}{3}z^2 r^{-1} & r\geq |z| \\[1ex]
   4\pi |z|+\frac{4\pi}{3}r^2 |z|^{-1} & r\leq |z|
   \end{array}\right. , \]
and thus by differentiation
\begin{equation}\label{formeliv}
   \int_{|\omega|=1}|z-r\omega|^{-1}(z-r\omega)\,d\omega
   =\left\{\begin{array}{c@{\quad:\quad}c}
   \frac{8\pi}{3r}\,z & r>|z| \\[1ex] 4\pi\bar{z}
   -\frac{4\pi}{3}r^2|z|^{-2}\bar{z} & r<|z|\end{array}\right. .
\end{equation}
Finally, for $z\in\R^3\setminus\{0\}$ also
\begin{equation}\label{formelv}
   \int |z-v|^{-1}|v|^{-3}v\,dv=2\pi\bar{z}
\end{equation}
can be computed.

\bigskip\bigskip

\noindent
{\bf Acknowledgements:} The authors are indebted to G.~Rein, A.~Rendall
and H.~Spohn for many discussions.

\end{document}